\documentclass[prd,letterpaper,twocolumn,aps,showpacs,nofootinbib,nobibnotes,superscriptaddress,preprintnumbers,floatfix]{revtex4-1}

\usepackage{amsfonts,amssymb}

\usepackage[figuresright]{rotating}
\usepackage[colorlinks=true,linkcolor=red,urlcolor=blue,citecolor=blue]{hyperref}
\usepackage{bm,color,amsmath}
\usepackage{mathrsfs}
\usepackage{xcolor}
\usepackage{dcolumn}
\usepackage{multirow}
\usepackage[normalem]{ulem}
\graphicspath{{./fig/}{./ps/}}

\newcommand{\bSOsig}{$0.47$}
\newcommand{\bSFOURsig}{$0.08$}
\newcommand{\bHDsig}{$0.036$}

\newcommand{\bSFOURsigTwo}{$0.16$}
\newcommand{\bHDsigTwo}{$0.072$}

\newcommand{\snrSO}{$0.2$}
\newcommand{\snrSFOUR}{$1.25$}
\newcommand{\snrHD}{$3$}

\newcommand{\bSOsigTrad}{$0.53$}
\newcommand{\bSFOURsigTrad}{$0.16$}
\newcommand{\bHDsigTrad}{$0.042$}

\newcommand{\stonybrook}{Physics and Astronomy Department, Stony Brook University, Stony Brook, New York 11794, USA}

\newcommand{\kipmu}{Kavli IPMU (WPI), UTIAS, University of Tokyo, Kashiwa, 277-8583, Japan}

\begin{document}

\title{Finding evidence for inflation and the origin of galactic magnetic fields \\with CMB surveys}

\author{Sayan Mandal}
\affiliation{\stonybrook}

\author{Neelima~Sehgal}
\affiliation{\stonybrook}

\author{Toshiya Namikawa}
\affiliation{\kipmu}

\begin{abstract}
The origin of the $\mu\mathrm{G}$ magnetic fields observed in galaxies is unknown.  One promising scenario is that magnetic fields generated during inflation, larger than 0.1 $\mathrm{nG}$ on Mpc scales, were adiabatically compressed to $\mu\mathrm{G}$ strengths in galaxies during structure formation.  Thus, detecting a scale-invariant primordial magnetic field (PMF) above $0.1\,\mathrm{nG}$ on Mpc scales just after recombination would indicate an inflationary origin of galactic magnetic fields. This would also provide compelling evidence that inflation occurred since only an inflationary mechanism could generate such a strong, scale-invariant magnetic field on Mpc scales.  In contrast, constraining the scale-invariant PMF strength to be below $0.1\,\mathrm{nG}$ would imply an inflationary scenario is not the primary origin, since such weak PMFs cannot be amplified enough via adiabatic compression to produce the strength of the galactic fields we observe today. We find that measurements of anisotropic birefringence by future CMB surveys will be able to improve the sensitivity to Mpc-scale inflationary PMFs by an order of magnitude, and, in particular, that CMB-HD would lower the upper bound to \bHDsigTwo~$\mathrm{nG}$ at the $95\%$ C.L., which is below the critical $0.1\,\mathrm{nG}$ threshold for ruling out a purely inflationary origin. If inflationary PMFs exist, we find that a CMB-HD survey would be able to detect them with about $3\sigma$ significance or higher, providing evidence for inflation itself.  
\end{abstract}

\maketitle

\section{Introduction}
\label{intro}

Magnetic fields observed in galaxies today are of order $\mu\mathrm{G}$, and their origin is an outstanding unsolved problem in astrophysics~\cite{Widrow:2002ud, Durrer:2013pga}.
It is believed that these magnetic fields can originate in either of three processes: (i) during inflation or related processes such as reheating or preheating, (ii) during early Universe phase transitions, e.g., the electroweak (EW) or quantum chromodynamics (QCD) phase transitions, or (iii) from weak seed fields or local plasma mechanisms in galaxies which are amplified by galactic dynamo processes~\cite{Kulsrud:2007an}.
The first two options above comprise the primordial origin scenario, and the magnetic fields generated in this way are called \textit{primordial magnetic fields} (PMFs)~\cite{Kandus:2010nw}.
In this scenario, Mpc-scale magnetic fields of order $\mathrm{nG}$ at the time of recombination are adiabatically compressed during the process of structure formation to $\mu\mathrm{G}$ strengths today in kpc-scale galaxies.  A feature in favor of the PMF scenario is that it can explain the observed presence of weak magnetic fields today, of at least $10^{-6}\,\mathrm{nG}$ on Mpc scales, outside of galaxies as well as in the empty voids between galaxy clusters~\cite{Neronov:2010gir, Fermi-LAT:2018jdy, VERITAS:2017gkr}.
Given that these weak fields are observed to be relatively uniform, irrespective of galaxy proximity, generating these fields with a galactic dynamo scenario is challenging.

PMFs can be generated during inflation from the amplification of quantum vacuum fluctuations.
Note that by inflation we are referring to any process that involves the shrinking of the comoving Hubble horizon before the onset of the standard cosmological phase of the Universe, and we are agnostic with regards to the microphysics that causes this phase which can be produced via a number of mechanisms~\cite{Starobinsky:1979ty, Guth:1980zm, Linde:1981mu, Albrecht:1982wi, Vilenkin:1983xq, Brandenberger:2000as, Armendariz-Picon:1999hyi, Brandenberger:2011eq, Brandenberger:2021kjo, Brandenberger:2021zib}.
The rapid stretching of the magnetic field due to the exponential expansion of the Universe is a natural way to explain the large correlation scales ($\sim$ Mpc) of the magnetic fields observed today~\cite{Widrow:2002ud, Ratra:1991bn}.
PMFs generated during inflation have a scale-invariant (or a nearly scale-invariant) spectrum.
To generate the magnetic fields observed today via the inflation scenario, the electromagnetic fields must be coupled to other fields in a way that slows down its energy density dilution during the inflationary expansion -- these can be couplings with the inflaton field responsible for inflation, some other scalar field in operation at that time, or the curvature of the spacetime (see Ref.~\cite{Kandus:2010nw} for a detailed review).

PMFs can also be generated from phase transitions in the early Universe, and there are two phase transitions of interest: the EW which occurred when the Universe had a temperature of about $T\sim 100\,\mathrm{GeV}$, and the QCD which happened around $T\sim 150\,\mathrm{MeV}$.
Magnetic fields generated via this process are causal, and their correlation length is limited by the Hubble radius at the time of generation.
In addition, the PMF generation mechanisms need to have very strong first order phase transitions that can generate significant turbulence in the primordial plasma, amplifying the fields~\cite{Hogan:1983zz, Witten:1984rs}.
Within the standard model of particle physics, however, these transitions are crossovers (i.e., smooth transitions) rather than being of the first-order (i.e., involving discontinuities)~\cite{Aoki:2006we, Kajantie:1996mn}.
Much research has been devoted to extending the standard model in a way that allows for first-order phase transitions~\cite{Grojean:2004xa, Huber:2006ma, Gorbunov:2011zzc}, but there is no evidence so far to favor these models.
Thus the inflationary scenario is a more natural way to produce PMFs.

Detecting scale-invariant PMFs above $\sim 0.1\,\mathrm{nG}$ on $\mathrm{Mpc}$ scales just after recombination would indicate that the seed magnetic fields responsible for the $\mu\mathrm{G}$ magnetic fields we see in galaxies have an {\it{inflationary}} origin~\cite{Grasso:2000wj, Widrow:2011hs, Ryu:2011hu}. The $\mu\mathrm{G}$ field would arise due to conservation of magnetic flux during adiabatic compression, which necessitates that the magnetic field increase inversely with the square of the size of the system~\cite{Grasso:2000wj}, i.e.$~0.1\,\mathrm{nG} =  1\,\mu\mathrm{G} (10~{\rm kpc} / 1~{\rm Mpc})^2$, where 10 kpc is a conservative lower limit for the radius of baryonic matter in galaxies. Moreover, PMFs from early Universe phase transitions would not be scale invariant. Such a detection 
would also be compelling evidence that inflation in fact occurred, since only an inflationary mechanism could generate such a strong, scale-invariant magnetic field on Mpc scales~\cite{Durrer:2013pga, Subramanian:2015lua}.  If scale-invariant PMFs are constrained to be below $0.1\,\mathrm{nG}$ on Mpc scales just after recombination, then the inflationary scenario can be ruled out as the primary source of magnetic fields in galaxies.  This is because such weak PMFs cannot be amplified enough during adiabatic compression to produce the strength of the fields we observe today.  

We can constrain the strength of PMFs on Mpc scales by looking for their imprints on the cosmic microwave background (CMB)~\cite{Pogosian:2018vfr}.  PMFs are imprinted on the CMB in at least two ways. (i)~PMFs induce perturbations in the spacetime metric, as well as a Lorentz force on the ions in the primordial plasma prior to recombination; these induce temperature and polarization anisotropies in the CMB~\cite{Durrer:2006pc, Subramanian:2006xs, Subramanian:2015lua, Paoletti:2008ck, Paoletti:2019pdi}, which can be probed by CMB power spectra measurements.
In addition, (ii)~PMFs present just after recombination rotate the polarization of CMB photons, causing anisotropic birefringence~\cite{Kosowsky:1996yc, Kosowsky:2004zh, Campanelli:2004pm}; this is also referred to as Faraday rotation in the literature.
Since the impact of PMFs on the CMB power spectra scales as the fourth power of the magnetic field strength, while the impact on anisotropic birefringence scales as the square of the magnetic field strength~\cite{Durrer:2013pga, Subramanian:2015lua},
as future CMB surveys improve their sensitivity, we expect to obtain tighter bounds on the PMF strength from CMB anisotropic birefringence, as opposed to CMB temperature and polarization spectra.

It has been suggested that PMFs present prior to recombination may generate small-scale baryonic clumping~\cite{Jedamzik:2011cu, Jedamzik:2018itu}, which may alter the large-scale CMB anisotropies and interestingly may relieve the Hubble tension~\cite{Jedamzik:2020krr, Rashkovetskyi:2021rwg, Galli:2021mxk}. We note that the prediction of baryon clumps due to the presence of PMFs depends on the details of the magnetohydrodynamic (MHD) simulation used~\cite{Jedamzik:2018itu}. 
Birefringence measurements provide a way to constrain PMFs in a simulation-independent way.

In this paper, we forecast constraints on the strength of a scale-invariant PMF from the measurement of anisotropic birefringence expected from the SO~\cite{SimonsObservatory:2018koc}, CMB-S4~\cite{Abazajian:2019eic}, and CMB-HD~\cite{Sehgal:2019ewc, Sehgal:2020yja} surveys.
We use a realistic power spectrum of inflationary PMFs that incorporates the effect of magnetohydrodynamic turbulence in the primordial plasma prior to recombination, and is scale-invariant at large scales.
This realistic PMF spectrum tightens constraints compared to the traditionally used spectrum of prior works~\cite{Kosowsky:2004zh, De:2013dra, Pogosian:2019jbt}.

Throughout this work, we use natural units, setting $\hbar=c=k_B=1$; we also set the permeability of free space to unity, i.e., $\mu_0=1$, and thereby express all electromagnetic quantities in Lorentz-Heaviside units.

\section{Anisotropic birefringence from PMFs}
\label{sec:Biref}

The presence of PMFs in the Universe induces anisotropic birefringence of the plane of polarization of the CMB~\cite{Kosowsky:1996yc, Kosowsky:2004zh, Campanelli:2004pm, Pogosian:2011qv}.
The anisotropic birefringence also converts $E$ modes of the CMB into $B$ modes, and induces cross correlations among the $E$, $B$, and $T$ modes~\cite{Harari:1996ac, Kahniashvili:2008hx, Pogosian:2011qv}.
The observed rotation angle $\alpha$ along a given direction, $\hat{\mathbf{n}}$, to the CMB is given in terms of the comoving magnetic field $\mathbf{B}(\mathbf{x})$ by
\begin{equation}\label{eAlphaDef}
\alpha(\hat{\mathbf{n}})=\frac{3}{16\pi^2\nu_0^2 q}\int_{\eta_{\rm dec}}^{\eta_0}d\eta\,\dot{\tau}(\eta)\,\mathbf{B}(\mathbf{x})\cdot\hat{\mathbf{n}},
\end{equation}
where $\mathbf{x}$ is the comoving position, $q$ is the electron charge, $\alpha_{\rm EM}=q^2$ is the fine-structure constant, $\nu_0$ is the frequency of observation, and $\dot{\tau}=x_e n_e \sigma_T a$ is the differential optical depth along the line of sight.
In addition, $a$ is the scale factor of the Universe, $\sigma_T=8\pi\alpha^2_{\rm EM}/3m_e^2$ is the Thomson scattering cross section, $m_e$ is the electron mass, and the inhomogeneities in $\dot{\tau}$ have been neglected.
Here, $\eta$ is the conformal time, defined in terms of the physical time $t$ as $d\eta=dt/a$, and the comoving magnetic field $\mathbf{B}$ is defined in terms of the physical magnetic field $\mathbf{B}_{\rm phys}$ as $\mathbf{B}=\mathbf{B}_{\rm phys} a^2$.
We note that the rotation angle depends on the observed frequency as $\alpha\propto\nu_0^{-2}$.
In Eq.~\eqref{eAlphaDef}, $\eta_{\rm dec}$ denotes the time of photon decoupling, while $\eta_0$ signifies the present time.

We assume the magnetic field $\mathbf{B}(\mathbf{x})$ to be a Gaussian random field in three dimensions.
Thus information about the energy and helicity of the magnetic field is encapsulated in the two-point correlation function.
We define $\mathbf{B}(\mathbf{k})=\int d^3x\,e^{i\mathbf{k}\cdot\mathbf{x}}\,\mathbf{B}(\mathbf{x})$, and write
\begin{equation}\label{ePMFPower}
    \begin{aligned}
    \left<B^*_i(\hat{\mathbf{k}})B_j(\hat{\mathbf{k}}')\right>&=(2\pi)^3 \delta^{(3)}(\mathbf{k}-\mathbf{k}')\left[(\delta_{ij}-\hat{k}_i\hat{k}_j)P_B(k)\right.\\
    &\left.i\epsilon_{ijl}\hat{k}_l P_H(k)\right],
    \end{aligned}
\end{equation}
where $P_B(k)$ and $P_H(k)$ are the symmetric and antisymmetric parts of the magnetic field power spectrum respectively.
We are interested in the former quantity $P_B(k)$, which is related to the energy density in the magnetic fields; the other term, $P_H(k)$, encodes the amount of magnetic helicity and does not contribute to birefringence~\cite{Kosowsky:2004zh, Pogosian:2011qv}.
The magnetic field power spectrum from PMFs is traditionally written as
\begin{equation}\label{eTradPMF}
P_B(k)=A_Bk^{n_B}, \quad k\leq k_D
\end{equation}
for some damping scale $k_D$, after which scale the spectrum cuts off because of small-scale viscous dissipation of the magnetic energy \cite{Kosowsky:2004zh, Durrer:2013pga}.
For inflationary PMFs, which are scale invariant on large scales, the PMF power-law index is $n_B=-3$.

We define $B_\lambda$ as the PMF strength smoothed over a region of comoving size $\lambda$ with a Gaussian kernel such that $B^2_\lambda=\int_0^\infty dk\,k^2\,e^{-k^2\lambda^2}\,P_B(k)/\pi^2$.
The PMF spectrum can then be expressed as
\begin{equation}\label{eTradPMFSpectraLamb}
P_B(k)=\frac{(2\pi)^{n_B+5}}{2}\frac{B^2_\lambda}{\Gamma\left(\frac{n_B+3}{2}\right)}\frac{k^{n_B}}{k_\lambda^{n_B}},\quad k\leq k_D,
\end{equation}
where $k_\lambda=2\pi/\lambda$ is the smoothing-scale wave number.
Based on the results of numerical simulations, the damping scale $k_D$ can be expressed as~\cite{Subramanian:1997gi, Jedamzik:1996wp, Mack:2001gc}
\begin{equation}\label{eCutoffScale}
\left(\frac{k_D}{\rm Mpc^{-1}}\right)^{n_B+5}\approx 2.9\times 10^4\left(\frac{B_\lambda}{10^{-9}\,\mathrm{G}}\right)^{-2} \left(\frac{k_\lambda}{\rm Mpc^{-1}}\right)^{n_B+3}.
\end{equation}
From this, the damping length scale, $1/k_D$, for 1 nG PMFs on Mpc scales, is much smaller than the Silk damping scale (i.e.~the thickness of the last scattering surface).
However, magnetic field components with characteristic wavelengths smaller than the thickness of the last scattering surface will generate polarization rotation at different optical depths; these rotations will nearly cancel one another, resulting in negligible net rotation~\cite{Kosowsky:2004zh}.
Thus, we set a value of $k_D=2\,\mathrm{Mpc}^{-1}$, which is approximately the Silk damping scale.
We set $\lambda$ to $1\,\mathrm{Mpc}$ since we are interested in constraining the PMF strength on these scales; we will henceforth refer to $B_{\lambda=1\,\mathrm{Mpc}}$ as $B$.

We follow Refs.~\cite{Kosowsky:2004zh, Kahniashvili:2008hx} to construct a two-point correlation function of the rotation angle, which can be expressed in terms of the PMF power spectrum as
\begin{equation}\label{eAlpha-TwoPt}
\begin{aligned}
\left<\alpha(\hat{\mathbf{n}})\alpha(\hat{\mathbf{n}}')\right>&\simeq\frac{9}{128\pi^5 q^2\nu_0^4}\sum_l\frac{2l+1}{4\pi}l(l+1)\,P_l(\hat{\mathbf{n}}\cdot\hat{\mathbf{n}}')\\
&\times\int dk\,k^2\,P_B(k)\left(\frac{j_l(k\eta_0)}{k\eta_0}\right)^2.
\end{aligned}
\end{equation}
The power spectrum $C_l^{\alpha\alpha}$ corresponding to this correlation function, defined as $\left<\alpha(\hat{\mathbf{n}})\alpha(\hat{\mathbf{n}}')\right>\equiv \sum_l(2l+1)C_l^{\alpha\alpha}\,P_l(\hat{\mathbf{n}}\cdot\hat{\mathbf{n}}')/4\pi$, can be written from Eqs.~\eqref{eTradPMFSpectraLamb} and~\eqref{eAlpha-TwoPt} as
\begin{equation}\label{eAnglePower}
\begin{aligned}
C_l^{\alpha\alpha}&\simeq\frac{9l(l+1)}{(4\pi)^3 q^2 \nu_0^4}\frac{B^2}{\Gamma\left(\frac{n_B+3}{2}\right)}\left(\frac{\mathrm{Mpc}}{\eta_0}\right)^{n_B+3}\\
&\times\int_0^{x_D} dx\,x^{n_B}\,j_l^2(x),
\end{aligned}
\end{equation}
where $x_D=k_D\eta_0$, and we assume $n_B$ is a constant over all $k$ scales of interest.
From $C_l^{\alpha\alpha}$ above, the amplitude of anisotropic birefringence, $A_\alpha$, is given by
\begin{equation}\label{eAmplitudeAlpha}
A_\alpha\equiv\frac{l(l+1)C_l^{\alpha\alpha}}{2\pi}\propto\frac{B^2}{\nu_0^4}.
\end{equation}
We highlight that the frequency dependence shown in Eq.~\ref{eAmplitudeAlpha} is unique to the birefringence signal caused by magnetic fields, helping to distinguish it from other sources of cosmic birefringence, including primordial gravitational waves~\cite{Zhao:2014rya}, Lorentz- and parity-violating physics~\cite{Carroll:1989vb}, and pseudoscalar fields coupled to electromagnetism~\cite{Carroll:1998zi, Pospelov:2008gg}.
We also note that since the PMF is a Gaussian random field with zero mean, i.e., $\langle\mathbf{B}(\mathbf{x})\rangle=0$, we can see from Eq.~\eqref{eAlphaDef} that the isotropic birefringence vanishes, i.e., $\langle\alpha(\hat{\mathbf{n}})\rangle=0$.

For a scale-invariant PMF spectrum, $A_\alpha$ will be independent of $l$ in the multipole region of interest. 
However, $A_\alpha$ depends on the frequency of observation.
Since the theoretically calculated $A_\alpha$ is frequency dependent, and CMB surveys generally observe at more than one frequency, we can construct an effective frequency, $\nu_{\rm eff}$, based on the frequency channels of the survey, at which to calculate the predicted $A_\alpha$.
Motivated by the $\nu_0^{-4}$ dependence of $A_\alpha$, we can construct an effective frequency $\nu_{\rm eff}$ for a two-frequency survey as
\begin{equation}\label{eEffFreq}
	\frac{1}{\nu_{\rm eff}^4}=\frac{1}{2}\left(\frac{1}{\nu_1^4}+\frac{1}{\nu_2^4}\right),
\end{equation}
where $\nu_1$ and $\nu_2$ are the two frequency channels for the given experiment.
This is equivalent to measuring $A_\alpha$ at the two frequency channels, and taking an arithmetic mean of the measurements, i.e.,
\begin{equation}
    A_{\alpha,{\rm measured}}=\frac{A_{\alpha,\nu_1}+A_{\alpha,\nu_2}}{2}\propto\frac{B^2}{\nu_1^4}+\frac{B^2}{\nu_2^4}\propto\frac{B^2}{\nu_{\rm eff}^4},
\end{equation}
under the assumption that the noise levels for the two frequency channels are equal.
Since CMB surveys usually have the most weight in the $90$ and $150\,\mathrm{GHz}$ channels, we find an effective frequency of $\nu_{\rm eff}=103.8\,\mathrm{GHz}$.

\begin{figure}[t]
    \begin{center}
    \includegraphics[width=\columnwidth]{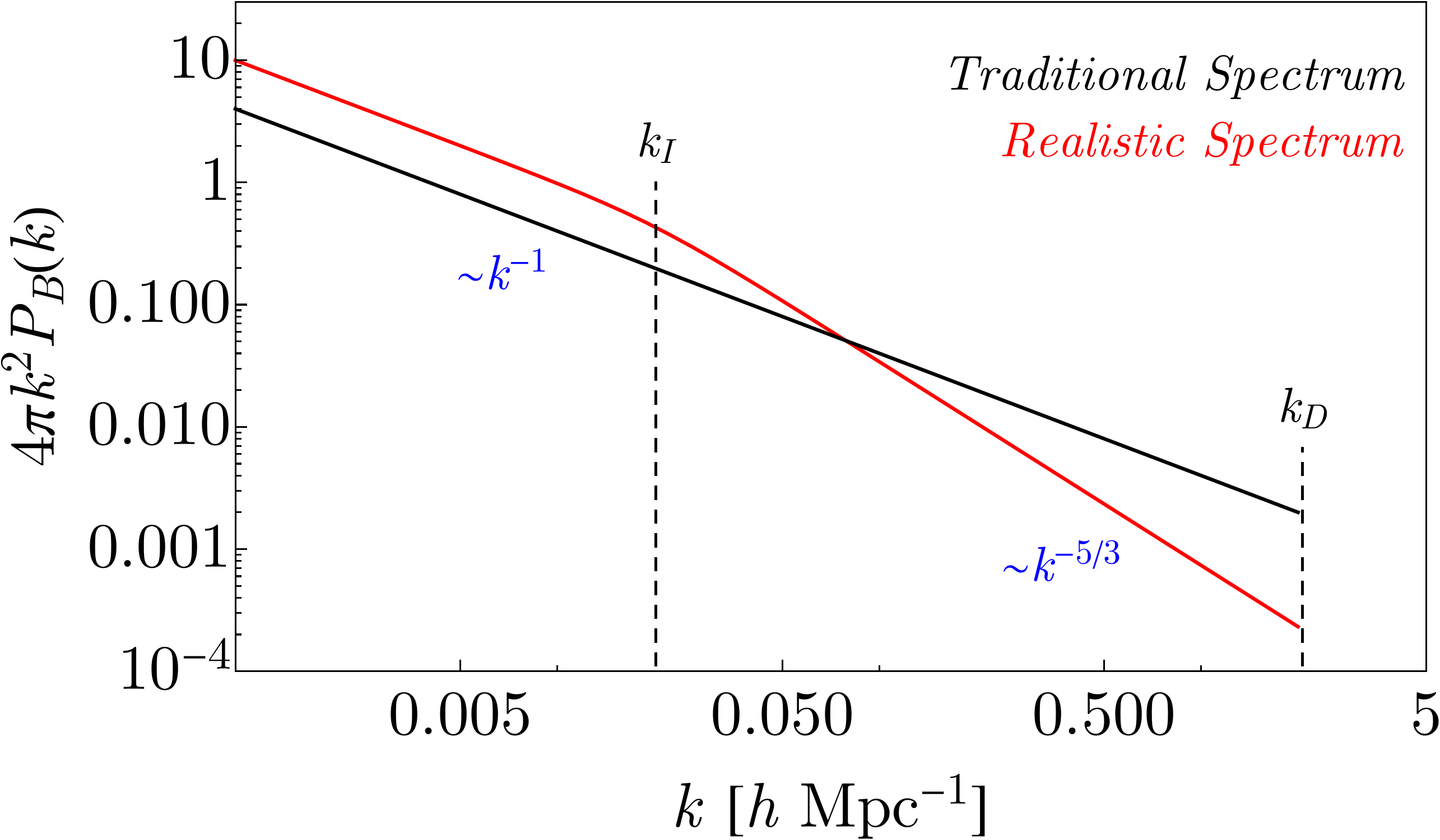}
    \end{center}
    \caption{The power spectrum $P_B(k)$ of inflationary PMFs used in calculating the rotation spectra of anisotropic birefringence. The black line shows the traditional spectrum of Eq. \eqref{eTradPMF}, while the red line shows the more realistic spectrum of Eq. \eqref{eRealPMF} used in this work. We have set $k_D=2\,\mathrm{Mpc}^{-1}$, and assumed a value of $k_I=k_D/100$~\cite{Brandenburg:2018ptt}. The $y$ axis is in dimensionless units where the energy spectrum is normalized with the initial energy density in magnetic fields.}
    \label{Fig:PMFConstr}
\end{figure}

\subsection{Effect of realistic PMF spectrum}

To account for the effect of MHD turbulence on PMFs, we modify the spectrum in Eq.~\eqref{eTradPMF} as
\begin{equation}\label{eRealPMF}
P_B(k)=\left\{ \begin{array}{l l}
A_1 k^{n_B} & \quad 0<k<k_I\\
A_2 k^{n_T} & \quad k_I<k<k_D
\end{array}, \right.
\end{equation}
where $n_B$ and $k_D$ are as above, and $n_T$ is the spectral index in the region $k_I<k<k_D$ where turbulence impacts the magnetic field.
Turbulence, in general, yields a Kolmogorov spectrum of $n_T=-11/3$.
Here $A_1$ and $A_2$ are related by the continuity condition $A_2=A_1 k_I^{n_B-n_T}$.
Figure~\ref{Fig:PMFConstr} shows the realistic spectrum in comparison to the traditional PMF spectrum in Eq.~\eqref{eTradPMF}.

In terms of the smoothed magnetic field $B\equiv B_{\lambda=1\,\mathrm{Mpc}}$, we can write the new rotation spectrum as
\begin{equation}\label{eAnglePowerReal}
\begin{aligned}
C_l^{\alpha\alpha}&\simeq\frac{9l(l+1)}{(4\pi)^3 q^2 \nu_0^4}\frac{B^2}{X}\left(\frac{\lambda}{\eta_0}\right)^3\left[\int_0^{x_I} dx\,\left(\frac{x}{x_I}\right)^{n_B}j_l^2(x)\right.\\
&\left.+\int_{x_I}^{x_D} dx\,\left(\frac{x}{x_I}\right)^{n_T}j_l^2(x)\right],
\end{aligned}
\end{equation}
with $x_D$ as before, $x_I=k_I\eta_0$, and $X$ defined in terms of the \textit{lower incomplete gamma function} $\gamma(s,y)$,
\begin{equation}\label{eXDef}
	\begin{aligned}
	X&=\frac{\gamma\left(\dfrac{n_B+3}{2},\lambda^2 k_I^2\right)}{\left(\lambda k_I\right)^{n_B}}\\
	&+\frac{\gamma\left(\dfrac{n_T+3}{2},\lambda^2 k_D^2\right)-\gamma\left(\dfrac{n_T+3}{2},\lambda^2 k_I^2\right)}{\left(\lambda k_I\right)^{n_T}}.
	\end{aligned}
\end{equation}
The realistic power spectrum of PMFs leads to a redistribution of magnetic energy at large scales, leading to a larger $C_l^{\alpha\alpha}$ compared to that for the traditional spectrum for a given $B$ strength, and consequently a tighter bound on $B$ from $A_\alpha$ measurements.

\section{Birefringence forecasts}
\label{sec:Methods}

\subsection{Reconstruction of anisotropic birefringence}

Anisotropic birefringence can be reconstructed by using the fact that the anisotropic polarization rotation angle, $\alpha(\hat{\mathbf{n}})$, mixes CMB $E$ and $B$ modes of different scales, leading to nonzero expectation values in the off-diagonal ($l,m\ne l',-m'$) elements of the CMB covariance.
Cross correlating different angular scales of $E$ and $B$ modes, we can reconstruct the anisotropies of the cosmic birefringence signal (see e.g. Refs.~\cite{Kamionkowski:2009:derot,Yadav:2012a,Namikawa:2016:rotsim}). 

Here, we describe how we compute the reconstruction noise spectrum of the anisotropic birefringence and constraints on $A_{\alpha}$.
We utilize an efficient algorithm to compute the curved sky reconstruction noise spectrum given by Eq.~(A6) of Ref.~\cite{Namikawa:2020:ACT-biref}, which requires the CMB polarization noise spectra.
For the CMB-HD experiment \cite{Sehgal:2020yja}, we assume a polarization white noise level of $0.7\,\mu$K$'$ which is close to the noise level when combining the $90$ and $150$\,GHz channels.
We assume a $0.4\,'$ Gaussian beam, observed sky fraction of $f_{\rm sky}=0.5$, and $90\%$ removal of the lensing $B$ modes. For comparison, we also consider the Simons Observatory (SO) \cite{SimonsObservatory:2018koc} and CMB-S4 \cite{Abazajian:2019eic}.
For SO, we consider the small aperture telescope and $3\,\mu$K$'$ white noise in polarization, $17\,'$ Gaussian beam, $f_{\rm sky}=0.1$, and $70\%$ removal of the lensing $B$ modes.
For CMB-S4, we consider the wide-area survey and assume a white noise in polarization of $2\,\mu$K$'$, a $2\,'$ Gaussian beam, $f_{\rm sky}=0.5$, and $85\%$ removal of the lensing $B$ modes.
We use CMB multipoles within $100\leq l\leq 5000$, since there is almost no improvement in the reconstruction noise level by further broadening this multipole range. The minimum multipole of the reconstructed birefringence spectrum is $l=2$. 

We compute $\sigma(A_\alpha)$ from (see e.g. Ref.~\cite{Pogosian:2019jbt})
\begin{equation}\label{sigmaA}
	\begin{aligned}
	\frac{1}{\sigma^2(A_\alpha)} 
	&= \sum_l f_{\rm sky}\frac{2l+1}{2}\frac{(C^{\alpha\alpha,\rm fid}_l)^2}{(N^{\alpha\alpha}_l)^2}
	\,,
	\end{aligned}
\end{equation}
where $C^{\alpha\alpha,\rm fid}_l=2\pi/l(l+1)$ and $N^{\alpha\alpha}_l$ is the reconstruction noise spectrum. Table~\ref{tab:TabBSIConstraints} shows our forecasted $\sigma(A_\alpha)$ constraints for each experiment. 
The current best upper bound on the scale-invariant anisotropic birefringence comes from ACTPol \cite{Namikawa:2020:ACT-biref} and SPTpol \cite{Bianchini:2020:SPT-biref} which roughly corresponds to $\sim 10$\,nG, while the tightest constraint on the PMF comes from the CMB power spectrum~\cite{Pogosian:2019jbt, Planck:2015zrl, SPT:2015htm}. However, we find that CMB-HD can improve the current bound on the anisotropic birefringence by roughly 4 orders of magnitude and will constrain the PMF much better than the CMB power spectrum in the future.

\begin{table}[t]
	\centering
	\begin{tabular}{|l||c|c|c|}
	    \cline{2-4}
		\multicolumn{1}{c|}{} & SO & CMB-S4 & CMB-HD \\
	    \hline\hline
		$\sigma(A_\alpha)$ ($\mathrm{deg}^2$) & $2.4\times 10^{-4}$ & $6.5\times 10^{-6}$ & $1.4\times 10^{-6}$ \\
		\hline
		$\sigma(B_{\rm{SI}})$ ($\mathrm{nG}$)& \bSOsig & \bSFOURsig & \bHDsig \\
		\hline
		SNR for $B_{\rm{SI}} = 0.1~\mathrm{nG}$ & \snrSO & \snrSFOUR & \snrHD \\
		\hline
	\end{tabular}
	\caption{Constraints on the amplitude of anisotropic birefringence, $A_\alpha$, and the corresponding scale-invariant PMF strength, $B_{\rm SI}$, obtained from expected measurements of future CMB surveys. 
	These constraints assume delensing (see text for details). 
	The $1\sigma$ uncertainties shown here assume a realistic PMF power spectrum. For a traditional PMF power spectrum, these 68\% C.L. constraints become \bSOsigTrad~nG, \bSFOURsigTrad~nG, and \bHDsigTrad~nG for SO, CMB-S4, and CMB-HD, respectively. The last row shows the detection significance (i.e. SNR = signal-to-noise ratio) for a scale-invariant PMF of strength $0.1$ nG for each experiment.}
	\label{tab:TabBSIConstraints}
\end{table}

\begin{figure*}[t]
    \begin{center}
    \includegraphics[width=0.9\textwidth]{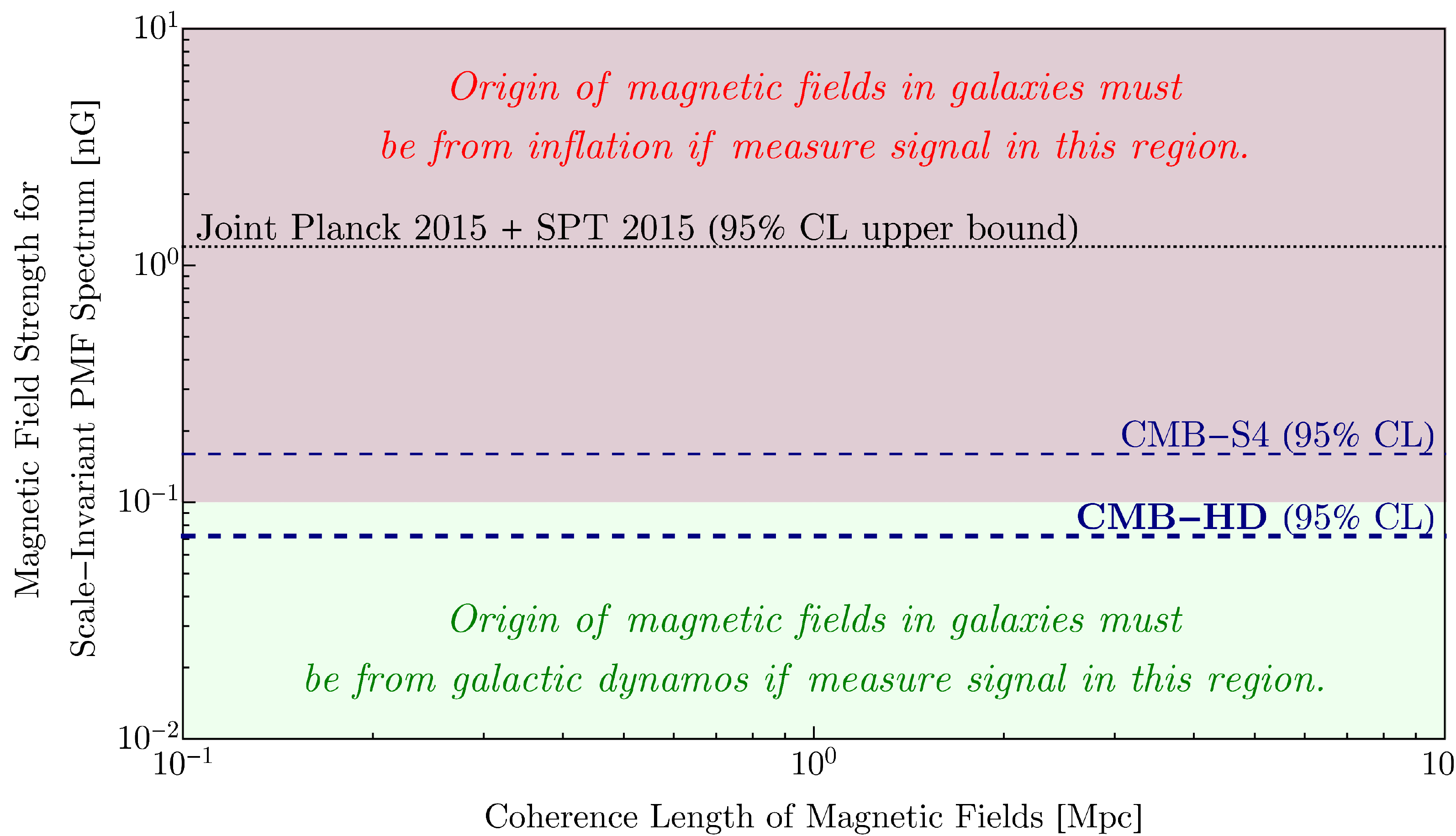}
    \end{center}
    \caption{Forecasted $95\%$ upper bounds on the strength of \textit{scale-invariant} PMFs, $B_{\rm SI}$, from measurements of the amplitude of anisotropic birefringence $A_\alpha$, for CMB-S4 (thin-dashed) and CMB-HD (thick-dashed lines). The dotted black line denotes the current $95\%$ C.L. upper bound on $B_{\rm SI}$ from a joint \textit{Planck} and SPT analysis of CMB temperature and polarization spectra~\cite{Planck:2015zrl, SPT:2015htm}. If $B_{\rm SI}$ is detected above $0.1\,\mathrm{nG}$, then the the origin of magnetic fields in galaxies must be from inflation; this would also be evidence for inflation itself. If $B_{\rm SI}$ is detected below $0.1\,\mathrm{nG}$, then the the origin of magnetic fields in galaxies must be from galactic dynamo processes. An upper limit below $0.1\,\mathrm{nG}$ would rule out an inflationary origin of galactic magnetic fields. A nondetection below $0.1\,\mathrm{nG}$ could still allow for galactic dynamo or electroweak phase transition mechanisms.}
    \label{Fig:FinalPlot}
\end{figure*}

\subsection{Subtracting birefringence from the Milky Way}

Magnetic fields in the Milky Way Galaxy can also lead to anisotropic birefringence of the CMB of order $A_\alpha\sim 10^{-5}\,{\rm deg^2}$, comparable to that expected from a $\mathcal{O}(0.1\,\mathrm{nG})$ PMF at Mpc scales~\cite{Oppermann:2011td, De:2013dra, Pogosian:2019jbt}.
Thus, in order to lower the bound on PMFs below $0.1\,\mathrm{nG}$, one would need to subtract the anisotropic birefringence due to the Milky Way.
To do this, one can use an independent measurement of the Galaxy-induced anisotropic birefringence obtained from measuring the polarization angle rotation, $\alpha(\hat{\mathbf{n}})$, of extragalactic radio sources distributed over the sky.
Since this rotation is frequency dependent, one can measure the polarization angle for each radio source at multiple frequencies to isolate the amount of angle rotation.
This has been done for about 40,000 extragalactic radio sources, taken from a compilation of catalogs including the NRAO VLA Sky Survey~\cite{Taylor_2009}.
From this, a map of the frequency-independent rotation measure, ${\rm RM}=\nu_0^2\alpha(\hat{\mathbf{n}})$ of the Milky Way was obtained, where $\alpha(\hat{\mathbf{n}})$ was measured for frequencies within the $\nu_0 \in (1,12)$~GHz range~\cite{Oppermann:2011td, De:2013dra}.
It was found that the RM spectrum for the Milky Way is nearly scale invariant, i.e., $l(l+1)C_l^{\rm RM}\propto l^{-0.17}$ for $l\lesssim 200$, where $A^2_{{\rm RM},l}\equiv l(l+1)C_l^{\rm RM}/2\pi \approx A^2_{\rm RM}$ and

\begin{equation}\label{eAAlphaARM}
    A_{\alpha}=2.363\times 10^{-7}\left(\frac{A_{\rm RM}}{1\,{\rm rad/m^2}}\right)^2\,{\rm deg^2},
\end{equation}
at an effective observation frequency of $\nu_{\rm eff}=103.8\,\mathrm{GHz}$.

The uncertainty in $A^2_{{\rm RM},l}$ arises both from sample variance and the uncertainty in the individual RM measurements of the radio sources.
Figure~3 of Ref.~\cite{De:2013dra} shows the RM spectrum induced by the Milky Way, and shows that for the cleanest 40\% of the sky, the Galactic RM contribution is roughly $A^2_{{\rm RM},l}\approx 70\,l^{-0.17}\,{(\rm rad/m^2)^2}$, with an associated error of roughly $\sigma_{A^2_{{\rm RM},l}}\approx 0.7\,A^2_{{\rm RM},l}$.
Since the amplitude $A_{\rm RM}$ is related to the magnitude of the scale-invariant magnetic field as $A_{\rm RM}= 68\,{\rm rad/m^2}\,(B_{\rm SI}/{1\,{\rm nG}})$, a measurement of $A_{RM}\approx \sqrt{70}\,{\rm rad/m^2}\approx 8\,{\rm rad/m^2}$ from the Galaxy corresponds to an effective Galactic magnetic field of $B_{\rm SI,G}\approx 0.12\,{\rm nG}$ on Mpc scales.

Given our estimate of $\sigma_{A^2_{{\rm RM},l}}$, and neglecting the covariance in the errors shown in Fig.~3 of Ref.~\cite{De:2013dra}, the signal-to-noise ratio for the detection of the scale-invariant Milky Way-induced $A^2_{\rm RM, G}$ is
\begin{equation}\label{eSIgToNoise}
    \left(\frac{S}{N}\right)^2=\sum_{l}\frac{\left(A^2_{{\rm RM},l}\right)^2}{\sigma^2_{A^2_{{\rm RM},l}}}=\sum_{l}\frac{\left(70\,l^{-0.17}\right)^2}{\left(0.7\times 70\,l^{-0.17}\right)^2}\approx 26^2.
\end{equation}
This value of the signal-to-noise ratio is likely optimistic, as it ignores the covariance of the errors bars. However, even if we conservatively consider a signal-to-noise ratio of 10, one can detect $A_{\rm RM, G}$ with an uncertainty of about $0.4~{\rm rad/m^2}$ and the effective Galactic magnetic field on Mpc scales with an uncertainty of $\sigma_{B_{\rm SI,G}}\approx 0.006\,{\rm nG}$.
This current uncertainty is subdominant to the $\sigma_{B_{\rm SI}}$ shown in Table~\ref{tab:TabBSIConstraints} from future CMB surveys, which suggests that anisotropic birefringence from the Galaxy can be subtracted.  Moreover, future CMB surveys will detect many more extragalactic radio sources, and the error on the Galactic rotation contribution will only improve.

In principle, non-Gaussian polarized synchrotron and dust emission from the Milky Way could also contaminate a measured birefringence spectrum.  This can be mitigated by removing multipoles below $l=100$, as is done for the forecasts here, as well as exploiting the full frequency range of upcoming CMB experiments. We leave detailed investigation of this to future work.

\section{Discussion}
\label{sec:Discuss}

In Fig.~\ref{Fig:FinalPlot}, we show our forecasted $95\%$ C.L. upper limits on the strength of scale-invariant PMFs, here denoted by $B_{\rm SI}$, from measurements of anisotropic birefringence for the CMB-S4 and CMB-HD experiments.
We also show the current $95\%$ C.L. upper bound on $B_{\rm SI}$ of $1.2\,\mathrm{nG}$ obtained from combining the TT, EE, and TE spectra from \textit{Planck} data and the BB spectrum from SPT~\cite{Pogosian:2019jbt}.
We see that anisotropic birefringence measurements from CMB-S4 will tighten the bound on $B$ to \bSFOURsigTwo~$\mathrm{nG}$ at $95\%$ C.L., while that from CMB-HD will tighten it further to \bHDsigTwo~$\mathrm{nG}$ at $95\%$ C.L. (see Table~\ref{tab:TabBSIConstraints}).
We note that the forecasted upper bound for CMB-HD lies below the $0.1\,\mathrm{nG}$ boundary that distinguishes between inflationary and dynamo origins of the galactic magnetic fields.  If PMFs at the Mpc scale are ruled out above $0.1\,\mathrm{nG}$, then we can rule out the inflationary scenario as the primary explanation for galactic magnetic fields.
Moreover, a {\it{detection}} of PMFs in the shaded green region of Fig.~\ref{Fig:FinalPlot} would suggest a galactic dynamo origin of galactic magnetic fields.
On the other hand, if $B_{\rm SI}$ is detected to be above $0.1\,\mathrm{nG}$, then it will be compelling evidence that galactic magnetic fields have an inflationary origin, and that inflation in fact occurred.  A CMB-HD survey would have the capability of detecting these inflationary PMFs at about the $3\sigma$ level or higher.

\vspace{3mm}

\begin{acknowledgments}
The authors thank Levon Pogosian and Tina Kahniashvili for useful discussions. S.M. and N.S. acknowledge support from NSF Grant number AST-1907657 and DOE Award number DE-SC0020441.
SM also acknowledges support from the Shota Rustaveli National Science Foundation (SRNSF) of Georgia (grant FR/18-1462). T.N. acknowledges support from JSPS KAKENHI Grant Number JP20H05859 and World Premier International Research Center Initiative (WPI), MEXT, Japan. 
\end{acknowledgments}

\bibliographystyle{apsrev4-1}
\bibliography{ref.bib}

\end{document}